\documentclass[11pt]{article}
\usepackage{amsmath}
\usepackage{amsfonts}
\usepackage{amssymb}
\usepackage{theorem}

\def\Th{\Theta}
\def\l{\lambda}
\def\p{\partial}
\newtheorem{prop}{Proposition}

\newcommand{\dbar}{\bar{\partial}}

\newcommand{\be}{\begin{equation}}
\newcommand{\ee}{\end{equation}}
\newcommand{\bea}{\begin{eqnarray}}
\newcommand{\eea}{\end{eqnarray}}
\newcommand{\beaa}{\begin{eqnarray*}}
\newcommand{\eeaa}{\end{eqnarray*}}

\newcommand{\nn}{\nonumber}
\renewcommand{\d}{\mathrm{d}}
\newcommand{\diverg}{\mathop{\mathrm{div}}\nolimits}
\begin{document}
\title{`Interpolating' differential reductions of multidimensional integrable
hierarchies}
\author{
L.V. Bogdanov\thanks
{L.D. Landau ITP RAS, 142432 Chernogolovka, Moscow region,  Russia; e-mail
leonid@landau.ac.ru}}
\maketitle
\begin{abstract}
We transfer the scheme of constructing differential reductions,
developed recently for the case of the Manakov-Santini hierarchy, to the
general multidimensional case. We consider in more detail the four-dimensional
case, connected with the second heavenly equation and its generalization 
proposed by Dunajski. We give a characterization of
differential reductions in terms of the Lax-Sato equations
as well as in the framework of the dressing method based
on nonlinear Riemann-Hilbert problem. 
\end{abstract}
\section{Introduction}
Recently \cite{LVB10}
we have constructed a class of reductions of the hierarchy associated
with the system introduced by Manakov and Santini \cite{MS06}
(see also \cite{MS07}, \cite{MS08}),
\bea
u_{xt} &=& u_{yy}+(uu_x)_x+v_xu_{xy}-u_{xx}v_y,
\nn\\
v_{xt} &=& v_{yy}+uv_{xx}+v_xv_{xy}-v_{xx}v_y.
\label{MSeq}
\eea
In \cite{Dun08} it was shown that a simple differential reduction 
\be
\alpha u= v_x 
\label{redint}
\ee
(where $\alpha$ is a constant) 
of the Manakov-Santini system (\ref{MSeq})
corresponds to the interpolating system, which was introduced in
\cite{Dun08} as "the most general symmetry reduction of the
second heavenly equation by a 
conformal Killing vector with a null self-dual derivative".
The term `interpolating' is connected with the fact that the system 
(\ref{MSeq}) with the reduction (\ref{redint}) `interpolates' between
the dKP equation, arising as $\alpha\rightarrow 0$, and the system considered
in \cite{Pavlov03} (see also \cite{Dun04,MS02,MS04}),
arising at $\alpha\rightarrow \infty$. 
Using Lax-Sato formulation  
of the Manakov-Santini hierarchy \cite{BDM07,LVB09}, 
in \cite{LVB10} we introduced a class of reductions, 
such that the zero order  reduction of this class corresponds to the
dKP hierarchy, and the first order reduction gives a hierarchy associated
with the interpolating system. We presented the generating equation and the 
Lax-Sato form of a reduced hierarchy for the interpolating system
and also for the reduction of an arbitrary order. 
Reduction of every order $k$ contains a parameter $\alpha$ and is
in some sence interpolating, providing the dKP hierarchy for
$\alpha\rightarrow 0$ and Gelfand-Dikii reduction of the order $k$
for the Manakov-Santini hierarchy at $\alpha\rightarrow \infty$.
It is important to note that Gelfand-Dikii reduction for the Manakov-Santini
hierarchy
does't imply the stationarity with respect to some higher time.
A characterization of the class of reductions in terms of the dressing data was given. 

In this work we transfer the construction developed in \cite{LVB10}
to the case of general (N+2)-dimensional hierarchy \cite{BDM07,LVB09}.
The Manakov-Santini hierarchy corresponds to $N=1$, this case is nongeneric, and there
are some new features in the general case. We pay special attention 
to the case $N=2$, which contains Pleba\'nski second heavenly equation 
\cite{Pleb} and its
generalization proposed by Dunajski \cite{Dun02}.

It is interesting to note a deep structural similarity of our construction
with the picture of differential reductions in the standard (dispersionful)
integrable case \cite{BKP,Zakharov1,ZM,BF}, 
though the technical tools
are essentially different.
\section{The hierarchy}
The Manakov-Santini hierarchy represents a special case 
corresponding to N=1 of the
general (N+2)-dimensional hierarchy defined by the generating relation
\bea
(J_0^{-1}\d \Psi^0\wedge \d \Psi^1\wedge \dots \wedge \d \Psi^N)_-=0,
\label{analyticity0}
\eea
where $\Psi^0,\dots, \Psi^N$ are the series
\bea
&&
\Psi^0=\lambda+\sum_{n=1}^\infty \Psi^0_n(\mathbf{t}^1,\dots,\mathbf{t}^N)\l^{-n},
\label{form0}
\\&&
\Psi^k=\sum_{n=0}^\infty t^k_n (\Psi^0)^{n}+
\sum_{n=1}^\infty \Psi^k_n(\mathbf{t}^1,\dots,\mathbf{t}^N)(\Psi^0)^{-n},
\label{formk}
\eea
$1\leqslant k\leqslant N$, $\mathbf{t}^k=(t^k_0,\dots,t^k_n,\dots)$,
$(\cdots)_-$ is a projection to negative powers of $\lambda$,
$J_0$ is a determinant of Jacobian matrix $J$,
\bea
J_0=\det J, \quad J_{ij}=\partial_i \Psi^j,\quad 0\leqslant i,j \leqslant N,
\label{J_0}
\eea
where $\partial_0=\frac{\partial}{\partial \lambda}$,
$\partial_k=\frac{\partial}{\partial x^k}$ ($1\leqslant k \leqslant N$),
$x^k=t^k_0$.

Generating relation (\ref{analyticity0}) is equivalent to the set
of Lax-Sato equations
\bea
&&
\partial^k_n\mathbf{\Psi}=\sum_{i=0}^N\left((J^{-1})_{ki} (\Psi^0)^n)\right)_+
{\partial_i}\mathbf{\Psi},\quad 0\leqslant n\leqslant \infty\, ,
1\leqslant k \leqslant N,
\label{genSato}
\eea
where $\mathbf{\Psi}=(\Psi^0,\dots,\Psi^N)$,
$(\cdots)_+$ is a projection to nonnegative powers of $\lambda$.
First flows of the hierarchy read
\bea
\partial^k_1\mathbf{\Psi}=(\lambda \partial_k-\sum_{p=1}^N (\partial_k u_p)\partial_p-
(\partial_k u_0)\partial_\lambda)\mathbf{\Psi},\quad 0<k\leqslant N,
\label{genlinear}
\eea
where $u_0=\Psi^0_1$,
$u_k=\Psi^k_1$, $1\leqslant k\leqslant N$.
A compatibility condition for any pair of linear equations  
(e.g., with $\partial^k_1$ and $\partial^q_1$, $k\neq q$)
implies closed nonlinear 
(N+2)-dimensional  system of PDEs for the set of functions $u_k$, $u_0$,
which can be written in the form
\bea
&&
\partial^k_1\p_q\hat u-\partial^q_1\p_k\hat u+[\p_k \hat u,\p_q \hat u]=
(\p_k u_0)\p_q-(\p_q u_0)\p_k,
\nn\\
&&
\partial^k_1\p_q u_0 - \partial^q_1\p_k u_0 + (\p_k \hat u)\p_q u_0 -
(\p_q \hat u)\p_k u_0=0,
\label{Gensystem}
\eea
where $\hat u$ is a vector field, $\hat u=\sum_{p=1}^N u_k \p_k$. 
\section{`Intepolating' differential reductions}
Lax-Sato equations of the hierarchy (\ref{genSato}) 
\bea
\partial^k_n\mathbf{\Psi}=\hat V^k_n\mathbf{\Psi}, 
\quad \hat V^k_n=\sum_{i=0}^N V^k_{n\,i}\p_i,\quad 
V^k_{n\,i}=\left((J_0^{-1})_{ki} (\Psi^0)^n)\right)_+
\label{genSato1}
\eea
imply
linear equations for the Jacobian (\ref{J_0})
\bea
\partial^k_n{J_0}=\sum_{i=0}^N\p_i\left(V^k_{n\,i}{J_0} \right).
\label{genSatoJ0}
\eea
These equations may be considered as formally adjoint to linear
equations (\ref{genSato1}), where we define adjoint operator
for $u_i\p_i$ as $-\p_i u_i$. It is interesting to note that the adjoint
equations coincide with equations (\ref{genSato1}) only in the case
of divergence-free vector fields.

Equivalently, equations (\ref{genSatoJ0}) can be written as
nonhomogeneous linear equations for the logarithm of Jacobian,
\bea
\partial^k_n\ln{J_0}=\hat V^k_n\ln{J_0}  + \diverg \hat V^k_n,\quad
\diverg \hat V^k_n=\sum_{i=0}^N\p_i V^k_{n\,i}.
\label{genSatoJ1}
\eea
The coefficients of these equations are polynomial in $\lambda$,
the function $(\ln J_0 - \alpha (\Psi^0)^k)$ is a solution of the equations,
thus the condition 
\bea
(\ln J_0 - \alpha (\Psi^0)^k)_-=0.
\label{red00}
\eea
defines a reduction of the hierarchy (this condition is
preserved by the dynamics). Similar to the case of the Manakov-Santini
hierarchy, this condition is characterized by  the existence 
of a polynomial solution of equations (\ref{genSatoJ1}).
\begin{prop}
The existence of a polynomial solution of the form
\be
f=-\alpha \l^k +\sum_0^{i=k-2} f_i(\mathbf{t})\l^i,
\label{f}
\ee
(where coefficients $f_i$ don't contain constants)
of equations (\ref{genSatoJ1}),
\beaa
\partial^k_n f=\hat V^k_n f  + \diverg \hat V^k_n,
\eeaa
is equivalent to the reduction condition (\ref{red00}).
\end{prop}
\textbf{Proof} The proof is completely analogous to the proof
of similar statement for the Manakov-Santini hierarchy, given in 
\cite{LVB10}.
First, the reduction condition (\ref{red00}) directly implies that
$f=(\ln J_0 - \alpha (\Psi^0)^k)$ is a polynomial solution 
of equations (\ref{genSatoJ1})
of required form, thus the existence of a polynomial solution is necessary.

To prove that it is sufficient, we note that $F=\ln J_0 -f$ solves homogeneous equations
(\ref{genSatoJ1}) (equations (\ref{genSato1})).
Let us expand $\l$ into the powers of $\Psi^0$, reverting the series (\ref{form0}),
and represent $F$ in the form
$$
F=\alpha (\Psi^0)^k + \sum_{-\infty}^{i=k-2} F_i(\mathbf{t})(\Psi^0)^i.
$$
It is easy to check that $F$ solves 
homogeneous equations (\ref{genSatoJ1}) 
iff all the coefficients $F_i(\mathbf{t})$ are constants.
Suggesting that the coefficients $f_i$ of the polynomial $f$ don't contain constants,
we come to the conclusion that $\ln J_0 -\alpha (\Psi^0)^k=f$,
thus $(\ln J_0 - \alpha (\Psi^0)^k)_-=0$ \hfill$\square$\\

Another equivalent formulation of the reduction can be presented in
terms of the generating relation (\ref{analyticity0}).
Reduction condition (\ref{red00}) implies that
$$
J_0=\exp(\alpha(\Psi^0)^k_-),
$$
and it is easy to prove the following statement:
\begin{prop}
The reduced hierarchy defined by the generating relation 
(\ref{analyticity0}) together with the reduction condition
(\ref{red00}) is equivalent to the generating relation
\bea
(\exp(-\alpha(\Psi^0)^k)
\d\Psi^0\wedge \d \Psi^1\wedge \dots \wedge \d \Psi^N)_-=0.
\label{analyticity0red}
\eea
\end{prop}

To calculate the reduction in terms of equations (\ref{Gensystem}),
it is convenient to start from nonhomogeneous linear equations
(\ref{genSatoJ1}), coresponding to the first flows of the hierarchy 
(\ref{genlinear}), which read
\bea
\partial^n_1 f=(\lambda \partial_n-\sum_{p=1}^N (\partial_n u_p)\partial_p-
(\partial_n u_0)\partial_\lambda)f - \p_n\sum_{p=1}^N \p_p u_p
, 1\leqslant n\leqslant N.
\label{genlinear1}
\eea
Substituting $f$ of the form (\ref{f}) to these equations, we 
obtain $k-1$ equations for $k$ coefficients $f_i$, which determine 
$f_i$ through the functions $u_0$, $u_n$ and define a differential reduction.

\subsubsection*{k=0. Divergence-free vector fields}
For $k=0$ the reduction condition (\ref{red00}), taking into
account that the expansion of $J_0$ is of the form $J_0=1+J_0^1\l^{-1}+\dots$
(so $(\ln J_0)_+=0$), implies that $J_0=1$. Thus nonhomogeneous linear equations
(\ref{genSatoJ1}) possess a solution equal to zero, and substituting it to the 
equations, we get
$$
\diverg \hat V^k_n=0,
$$
so the vector fields in the Lax-Sato equations (\ref{genSato1})
are divergence-free, and  
the flows
of the hierarchy are volume-preserving. 
Generating equation (\ref{analyticity0}) for the reduced hierarchy
reads
\beaa
(\d \Psi^0\wedge \d \Psi^1\wedge \dots \wedge \d \Psi^N)_-=0.
\eeaa

On the other hand, volume-preserving reduction can be obtained from
the reduction (\ref{red00}) with arbitrary $k$ in the limit 
$\alpha\rightarrow 0$. Thus the reduction (\ref{red00}) with arbitrary $k$
is an `interpolating' reduction between the volume-preserving hierarchy
and the hierarchy, characterized by the existence of polynomial solution
of Lax-Sato equations (\ref{genSato}) (Gelfand-Dikii reduction).

To calculate the reduction in terms of equations (\ref{Gensystem}),
we substitute zero solution to equations (\ref{genlinear1}) and obtain a
condition
\beaa
\diverg \hat u:=\sum_{p=1}^N \p_p u_p=0.
\eeaa
Thus equations (\ref{Gensystem}) reduce to divergence-free 
vector fields $\hat u$.
\subsubsection*{The reduction for $\mathbf{k=1}$}
For $k=1$ the reduction condition (\ref{red00}) reads
\beaa
(\ln J_0 - \alpha \Psi^0)_-=0,
\eeaa
thus
\beaa
\ln J_0 - \alpha \Psi^0=-(\alpha\Psi^0)_+ ,
\eeaa
and for the Jacobian we obtain
\be
J_0=\exp\alpha(\Psi^0-\lambda).
\label{redk1}
\ee
Generating equation for the reduced hierarchy is
\beaa
(\exp(-\alpha\Psi^0)\d \Psi^0\wedge \d \Psi^1\wedge \dots \wedge \d \Psi^N)_-=0.
\eeaa

The reduction implies the 
existence of the solution $-\alpha\lambda$ of nonhomogeneous 
linear equations (\ref{genSatoJ1})
and leads to the relations
\beaa
\diverg \hat V^k_n=\alpha V^k_{n\,0}.
\eeaa
Using these relations for the first flows (\ref{genlinear1}), we
calculate the reduction condition for equations (\ref{Gensystem}),
\beaa
\diverg \hat u:=\sum_{p=1}^N \p_p u_p= \alpha u_0.
\eeaa
Using this relation, it is possible to exclude the function $u_0$
from the system (\ref{Gensystem}) and obtain the reduced system in
explicit form,
\bea
&&
\partial^k_1\p_q\hat u-\partial^q_1\p_k\hat u+[\p_k \hat u,\p_q \hat u]=
\alpha^{-1}((\p_k\diverg \hat u)\p_q-(\p_q \diverg \hat u)\p_k).
\label{Gensystemk1}
\eea
The limit $\alpha\rightarrow 0$ of the reduced hierarchy corresponds
to the volume-preserving case (k=0) described above, while the limit
$\alpha\rightarrow \infty$ corresponds to the hierarchy characterized by the
relation $\Psi^0=\l$ \cite{BK},\cite{BDM07}. 
For this hierarchy vector fields of Lax-Sato equations
(\ref{genSato}) do not contain a derivative with respect to a spectral variable,
and $u_0$ in equations (\ref{Gensystem}) is equal to zero,
\bea
&&
\partial^k_1\p_q\hat u-\partial^q_1\p_k\hat u+[\p_k \hat u,\p_q \hat u]=0.
\label{Gensystemk10}
\eea
This
hierarchy is a `precursor' of hyper-Kahler hierarchies 
\cite{Takasaki,Takasaki1}, 
which correspond
to Hamiltonian vector fiels both in Lax-Sato equations (\ref{genSato})
and in equations (\ref{Gensystem}). Thus the reduction
(\ref{red00}) with $k=1$
is `interpolating' between the volume-preserving hierarchy,
connected with the system (\ref{Gensystem}) for divergence-free vector 
fields $\hat u$,
and pre-hyper-Kahler
hierarchy connected with the system (\ref{Gensystemk10}).
\subsubsection*{$\mathbf{k=2}$}
For $k=2$ the reduction condition (\ref{red00}) reads
\beaa
(\ln J_0 - \alpha (\Psi^0)^2)_-=0,
\eeaa
thus
\beaa
\ln J_0 - \alpha (\Psi^0)^2=-\alpha(\Psi^0)^2_+ ,
\eeaa
and for the Jacobian we obtain
\be
J_0=\exp\alpha((\Psi^0)^2_-).
\label{redk2}
\ee
Generating equation for the reduced hierarchy is
\beaa
(\exp(-\alpha(\Psi^0)^2)
\d \Psi^0\wedge \d \Psi^1\wedge \dots \wedge \d \Psi^N)_-=0.
\eeaa

The reduction implies the 
existence of polynomial solution $f=-\alpha\l^2 +f_1$ of nonhomogeneous 
linear equations (\ref{genSatoJ1}). Substituting this solution
to equations (\ref{genlinear1}),
we obtain
\beaa
&&
\partial^n_1 f_1=-(\partial_n \hat u) f_1 -
\p_n\diverg \hat u,\\
&&
\partial_n f_1=-2\alpha(\partial_n u_0),
\eeaa
or, excluding $f_1$,
\beaa
\partial^n_1 u_0+ (\partial_n \hat u)u_0 +
\frac{1}{2\alpha}\p_n\diverg \hat u=0.
\eeaa
These relations represent a differential reduction for the system
(\ref{Gensystem}). It is not difficult to check that 
the substitution of these relations to the second equation of the system 
(\ref{Gensystem}) satisfies it identically.

The limit $\alpha\rightarrow 0$ of the reduced hierarchy corresponds
to the volume-preserving case (k=0), while the limit
$\alpha\rightarrow \infty$ corresponds to the hierarchy characterized by the
relation $(\Psi^0)^2_-=0$, or, equivalently, $(\Psi^0)^2=\l^2+2u_0$ 
(Gelfand-Dikii reduction).
\subsubsection*{$\mathbf{k=3}$}
For $k=3$ the reduction condition (\ref{red00}) reads
\beaa
(\ln J_0 - \alpha (\Psi^0)^3)_-=0,
\eeaa
and for the Jacobian we obtain
\be
J_0=\exp\alpha((\Psi^0)^3_-).
\label{redk3}
\ee
Generating equation for the reduced hierarchy is
\beaa
(\exp(-\alpha(\Psi^0)^3)
\d \Psi^0\wedge \d \Psi^1\wedge \dots \wedge \d \Psi^N)_-=0.
\eeaa
The reduction implies the 
existence of polynomial solution $f=-\alpha\l^3+ f_1\l + f_2$ of nonhomogeneous 
linear equations (\ref{genSatoJ1}). Substituting this solution
to equations (\ref{genlinear1}),
we obtain
\beaa
&&
\partial^n_1 f_2=-(\partial_n \hat u)f_2 
+3\alpha u_0(\partial_n u_0)
-\p_n\diverg \hat u,\\
&&
\partial_n f_2=-{3\alpha}
(\partial_n \hat u)u_0
\eeaa
or, excluding $f_2$,
\beaa
\partial^n_1
\left(
(\partial_n \hat u)u_0
\right)=-
\partial_n 
\left(
(\partial_n \hat u) 
(\partial_n \hat u) u_0 
+u_0(\partial_n u_0)
-\frac{1}{3\alpha}\p_n\diverg \hat u
\right).
\eeaa
This relation represents a differential reduction for the system
(\ref{Gensystem}).

The limit $\alpha\rightarrow 0$ of the reduced hierarchy corresponds
to the volume-preserving case (k=0), while the limit
$\alpha\rightarrow \infty$ corresponds to the hierarchy characterized by the
relation $(\Psi^0)^3_-=0$
(Gelfand-Dikii reduction).
\section*{N=2. Systems connected with Plebanski second heavenly equation}
We will consider in more detail the construction of reductions in the
case $N=2$, which contains Plebanski second heavenly equation and its
generalization proposed by Dunajski \cite{Dun04}.
\subsubsection*{k=0} 
Volume-preserving reduction 
(in this case it is area-preserving) corresponds 
to Plebanski generalization
of the second heavenly equation \cite{BDM07}. Vector fields
in the Lax-Sato equations of the hierarchy (\ref{genSato1}) are
Hamiltonian (two-dimensional divergence-free), and it is possible
to write the reduced system (\ref{Gensystem}) in terms of the potential
$\Theta$, $u_1=\Theta_y$, $u_2=-\Theta_x$, $x=x^1$, $y=x^2$. After the identification
$z=-t^1_1$, $w=t^2_1$, $\phi=u_0$ we get the Dunajski system \cite{Dun04}
\bea
&&
\Th_{wx}+\Th_{zy}+\Th_{xx}\Th_{yy}-\Th_{xy}^2=\phi,
\nn\\
&&
\label{Dun}
\phi_{xw}+\phi_{yz}+
\Th_{yy}\phi_{xx}+\Th_{xx}\phi_{yy}-2\Th_{xy}\phi_{xy}=0.
\eea
The hierarchy connected with this system is studied in detail in \cite{BDM07}.
The general system (\ref{Gensystem}) in this notations reads
\bea
&&
(\partial_{zy}+\partial_{wx})\hat u+[\p_y \hat u,\p_x \hat u]=
(\p_y \phi)\p_x-(\p_x \phi)\p_y,
\nn\\
&&
(\partial_{zy}+\partial_{wx}+ (\p_y \hat u)\p_x -
(\p_x \hat u)\p_y )\phi=0,
\label{GensystemD}
\eea
where $\hat u=u_1\p_x+u_2\p_y$.
\subsubsection*{k=1}
The reduction with $k=1$ is characterized by the relation
\beaa
J_0=\exp\alpha(\Psi^0-\lambda).
\eeaa
Generating equation for the reduced hierarchy is
\beaa
(\exp(-\alpha\Psi^0)\d \Psi^0\wedge \d \Psi^1\wedge\d \Psi^2)_-=0.
\eeaa
The reduction implies the 
existence of the solution $-\alpha\lambda$ of nonhomogeneous 
linear equations (\ref{genSatoJ1}).
The reduction condition for equations (\ref{GensystemD})is
\beaa
\diverg \hat u:=\p_x u_1+\p_y u_2= \alpha \phi.
\eeaa
Using this relation, it is possible to exclude the function $u_0$
from the system (\ref{GensystemD}) and obtain the reduced system in
explicit form,
\beaa
&&
(\partial_{zy}+\partial_{wx})\hat u+[\p_y \hat u,\p_x \hat u]=
\alpha^{-1}((\p_y \diverg \hat u)\p_x-(\p_x \diverg \hat u)\p_y),
\eeaa
The limit $\alpha\rightarrow 0$ of the reduced hierarchy corresponds
to the Dunajski system hierarchy, while the limit
$\alpha\rightarrow \infty$ corresponds to the hierarchy characterized by the
relation $\Psi^0=\l$ \cite{BDM07,BK}. 
For this hierarchy vector fields of Lax-Sato equations
(\ref{genSato}) do not contain a derivative with respect to a spectral variable,
and $\phi$ in equations (\ref{GensystemD}) is equal to zero,
\bea
&&
(\partial_{zy}+\partial_{wx})\hat u+[\p_y \hat u,\p_x \hat u]=0,
\label{preheav}
\eea
This
hierarchy is a `precursor' of Plebanski second heavenly
equation hierarchy \cite{Takasaki,BK} corresponding
to Hamiltonian vector fiels both in Lax-Sato equations (\ref{genSato})
and in equation (\ref{preheav}),
which reduces to Plebanski second heavenly equation
\beaa
\Th_{wx}+\Th_{zy}+\Th_{xx}\Th_{yy}-\Th_{xy}^2=0.
\eeaa
Thus the reduction
(\ref{red00}) with $k=1$
is `interpolating' between the hierarchy,
connected with Dunajski system (\ref{Dun})
and the
hierarchy connected with the system (\ref{preheav}), which for
Hamiltonian vector fields reduces to Plebanski second heavenly
equation.
\subsubsection*{k=2}
Reduction with $k=2$ is characterized by the relation
\beaa
J_0=\exp(\alpha(\Psi^0)^2_-).
\eeaa
Generating equation for the reduced hierarchy is
\beaa
(\exp(-\alpha(\Psi^0)^2)\d \Psi^0\wedge \d \Psi^1\wedge\d \Psi^2)_-=0.
\eeaa
Reduction conditions in terms of the system (\ref{GensystemD})
are
\beaa
\partial_z \phi-(\partial_x \hat u) \phi -
\frac{1}{2\alpha}\p_x\diverg \hat u=0,
\nn\\
\partial_w \phi+(\partial_y \hat u) \phi +
\frac{1}{2\alpha}\p_y\diverg \hat u=0.
\eeaa
The limit $\alpha\rightarrow 0$ corresponds to Dunajski system (\ref{Dun}),
and the limit $\alpha\rightarrow \infty$ -- to the second Gelfand-Dikii reduction
$(\Psi^0)^2_-=0$ for the system (\ref{preheav}).
\subsubsection*{k=3}
Reduction with $k=3$ is characterized by the relation
\beaa
J_0=\exp(\alpha(\Psi^0)^3_-).
\eeaa
Generating equation for the reduced hierarchy is
\beaa
(\exp(-\alpha(\Psi^0)^3)\d \Psi^0\wedge \d \Psi^1\wedge\d \Psi^2)_-=0.
\eeaa
Reduction conditions in terms of the system (\ref{GensystemD})
are
\beaa
&&
\partial_z
\left(
(\partial_x \hat u)\phi
\right)=
\partial_x 
\left(
(\partial_x \hat u) 
(\partial_x \hat u) \phi 
+\phi(\partial_x \phi)
-\frac{1}{3\alpha}\p_x\diverg \hat u
\right),
\\
&&
\partial_w
\left(
(\partial_y \hat u)\phi
\right)=-
\partial_y 
\left(
(\partial_y \hat u) 
(\partial_y \hat u) \phi 
+\phi(\partial_y \phi)
-\frac{1}{3\alpha}\p_y\diverg \hat u
\right).
\eeaa
The limit $\alpha\rightarrow 0$ corresponds to Dunajski system (\ref{Dun}),
and the limit $\alpha\rightarrow \infty$ -- to the third Gelfand-Dikii reduction
$(\Psi^0)^3_-=0$ for the system (\ref{preheav}).
\section{Characterization of reductions in terms of the dressing data}
A dressing scheme for the hierarchy (\ref{analyticity0},\ref{genSato})
can be formulated
in terms of (N+1)-component nonlinear 
Riemann-Hilbert problem on the unit circle $S$
in the complex plane of the variable $\l$,
\bea
&&
\Psi^0_\text{in}=F_0(\Psi^0_\text{out},\Psi^1_\text{out},\dots,\Psi^N_\text{out}),
\nn\\
&&
\Psi^k_\text{in}=F_k(\Psi^0_\text{out},\Psi^1_\text{out},\dots,\Psi^N_\text{out}),
\quad 1\leqslant k\leqslant N,
\label{RiemannMS}
\eea
where the functions 
$\Psi^0_\text{in}(\l,\mathbf{t})$, $\Psi^k_\text{in}(\l,\mathbf{t})$ 
are analytic inside the unit circle,
the functions $\Psi^0_\text{out}(\l,\mathbf{t})$, $\Psi^k_\text{out}(p,\mathbf{t})$ 
are analytic outside the
unit circle and have an expansion of the form (\ref{form0}), (\ref{formk}).
The functions $F_0$, $F_k$ are suggested to define (at least locally) a
diffeomorphism in $\mathbb{C}^{N+1}$, 
$\mathbf{F}\in\text{Diff(N+1)}$, and we call them
the dressing data. In compact form the problem (\ref{RiemannMS}) can be written as
\be
\mathbf{\Psi}_\text{in}=\mathbf{F}(\mathbf{\Psi}_\text{out}).
\label{RiemannMSbis}
\ee
It is straightforward to demonstrate that the problem
(\ref{RiemannMS}) implies the analyticity of the differential form
$$
\Omega_0=J_0^{-1}\d \Psi^0\wedge \d \Psi^1\wedge \dots \wedge \d \Psi^N
$$
(where the independent variables of the
differential include all the times $\mathbf{t}$ and $\l$)
in the complex plane and the generating relation (\ref{analyticity0}), 
thus defining
a solution of the hierarchy. Considering a reduction to the group of
volume-preserving
diffeomorphisms \text{SDiff(N+1)}, we obtain a reduction of the general
hierarchy (\ref{analyticity0}) to the case $J_0=1$ 
(divergence-free vector fields), 
$$
(\d \Psi^0\wedge \d \Psi^1\wedge \dots \wedge \d \Psi^N)_-=0.
$$
For $N=2$ it is
the Dunajski system hierarchy. 

To construct a class of reductions (\ref{red00}), it is necessary
to consider a kind of `twisted' volume-preservation condition.
Let the functions $G_q(y_0,\dots,y_n)$,  $0\leqslant q\leqslant N$,
define a volume-preserving
diffeomorphism, $\mathbf{G}\in\text{SDiff(N+1)}$, 
$$
\left|\frac{D(G_0,\dots,G_N)}{D(y_0,\dots,y_N)}\right|=1,
$$
where for the Jacobian we use a notation 
\beaa
\left|\frac{D(y_1,\dots,y_N)}{D(x_1,\dots,x_N)}\right|=
\det \frac{D(y_1,\dots,y_N)}{D(x_1,\dots,x_N)}=
\det\left(\frac{\p y_i}{\p x_j}\right).
\eeaa

Let us fix a set of analytic functions $f_q(y_0,\dots,y_n)$
(the reduction data) defining a diffeomorphism
and consider a problem
\bea
\mathbf{f}(\mathbf{\Psi}_\text{in})=\mathbf{G}(\mathbf{f}(\mathbf{\Psi}_\text{out})),
\label{Riemannred}
\eea
which corresponds to the reduction of the hierarchy.
The reduction condition for the dressing data of the problem 
(\ref{RiemannMSbis}) reads
\be
\mathbf{f}\circ\mathbf{F}\circ\mathbf{f}^{-1}\in \text{SDiff(N+1)}.
\label{redMSdressing}
\ee
This is what we call a `twisted' volume-preservation condition.
In terms of equations of the
MS hierarchy the reduction is characterized by the relation
$$
(\d f_0(\mathbf{\Psi})\wedge \dots\wedge \d f_N(\mathbf{\Psi}))_\text{in}=
(\d f_0(\mathbf{\Psi})\wedge \dots\wedge \d f_N(\mathbf{\Psi}))_\text{out},
$$
thus the differential form 
$$
\Omega_\text{red}=\d f_0(\mathbf{\Psi})\wedge \dots\wedge \d f_N(\mathbf{\Psi})
$$
is analytic in the complex plane, and the reduced hierarchy is defined by
the generating relation
$$
(\d f_0(\mathbf{\Psi})\wedge \dots\wedge \d f_N(\mathbf{\Psi}))_-=0,
$$
or, equivalently,
$$
\left(\left|\frac{D(f_0,\dots,f_N)}{D(\Psi^0,\dots,\Psi^N)}\right|   
\d \Psi^0\wedge \d \Psi^1\wedge \dots \wedge \d \Psi^N
\right)_-=0.
$$
The diffeomorphism $\mathbf{f}$ for the reduced hierarchy is defined
modulo a volume-preserving diffeomorphism. To obtain the reduced hierarchy
(\ref{analyticity0red}), it is possible to use a symmetric choice
\bea
&&
f_0(\mathbf{\Psi})=\Psi^0,
\nn\\&&
f_n(\mathbf{\Psi})=
\exp(-\alpha N^{-1}(\Psi^0)^k)\Psi^n,\quad 1\leqslant n\leqslant N.
\label{reddiffMS}
\eea
Thus we come to the following conclusion:
\begin{prop}
In terms of 
the dressing data for the
problem (\ref{RiemannMSbis}), 
a class of reductions (\ref{red00})  is characterized 
by the condition (\ref{redMSdressing}), where $\mathbf{f}$ is defined by 
(\ref{reddiffMS}). 
\end{prop}
\section*{Acknowledgments}
The author is grateful to S.V. Manakov for
useful discussions.
This research was partially supported by the Russian Foundation for
Basic Research under grants no. 09-01-92439, 10-01-00787
and by the President of Russia
grant 4887.2008.2 (scientific schools). The participation in
the conference "Nonlinear Physics. Theory and experiment. VI",
where the talk resulting in this paper was presented, was made possible
due to RFBR travel grant 10-01-08119 and the hospitality of the
organizers.

\end{document}